\documentclass[twoside]{article}
\usepackage{fleqn,espcrc2,epsf}

\title{FermiQCD: A tool kit for parallel lattice QCD applications \\
{\tt [ http://latticeqcd.fnal.gov/software/fermiqcd/ ]}}

\author{Massimo Di Pierro\address[FNAL]{
Fermilab, Kirk and Pine St., Batavia, Illinois 60563, USA}%
        \thanks{Poster presented at Lattice 2001, Berlin}}       

\begin{document}

\begin{abstract}
We present here the most recent version of FermiQCD, a collection of C++ 
classes, functions and parallel algorithms for lattice QCD, based on Matrix 
Distributed Processing.
FermiQCD allows fast development of parallel lattice applications and 
includes some SSE2 optimizations for clusters of Pentium 4 PCs.
\vspace{1pc}
\end{abstract}

\maketitle

\section{Introduction}

FermiQCD
 is a collection of
classes, functions and parallel algorithms for lattice QCD~\cite{lattice}, 
written in C++. It is based on Matrix Distributed Processing\footnote{\tt 
http://www.phoenixcollective.org/mdp} (MDP)~\cite{mdp}.
The latter is a library that includes C++ methods for matrix manipulation, 
advanced statistical analysis (such as Jackknife and Boostrap) and
optimized algorithms for interprocess communications of distributed 
lattices and fields. These communications are implemented using
Message Passing Interface (MPI) but MPI calls are hidden to the high 
level algorithms that constitute FermiQCD.

FermiQCD works also on single processor computers and, in this case,
MPI is not required.

\section{FermiQCD overview}

The basic fields defined in FermiQCD are:
\vskip 3mm
\noindent {\bf class {\tt gauge\_field}:}

List of implemented algorithms:
\begin{itemize}
\item heatbath algorithm
\item anisotropic heatbath
\item $O(a^2)$ heatbath
\end{itemize}
These algorithms work for arbitrary gauge groups $SU(N_c)$, for 
arbitrary lattice dimensions and topologies.
FermiQCD also supports arbitrarily twisted boundary conditions for large 
$\beta$ computations and studies of topology.

\vskip 3mm
\noindent{\bf class {\tt fermi\_field}:}

List of implemented algorithms:
\begin{itemize}
\item multiplication by $Q=(D\!\!\!\!/+m)$, for Wilson and Clover actions,
for isotropic and anisotropic lattices
\item minimal residue inversion for $Q$
\item stabilized biconjugate gradient (BiCGStab) inversion for $Q$
\item Wupperthal smearing for the field
\item stochastic propagators
\end{itemize}
These algorithms work for arbitrary gauge groups $SU(N_c)$ and for 
arbitrary topologies in 4 dimensions.
The multiplication by $Q$, clover (isotropic and anisotropic), 
for $SU(3)$, is optimized using Pentium 4 
SSE2 instructions in assembler language. This  implementation is 
based on the assembler 
macro functions written by Martin L\"uscher~\cite{luscher}
\vskip 3mm
\noindent{\bf class {\tt fermi\_propagator}:}

This is an implementation of ordinary quark propagators. 
A {\tt fermi\_propagator} can be generated using any of the
inversion algorithms of a {\tt fermi\_field}.
\vskip 3mm

\noindent{\bf class {\tt staggered\_field}:}

Kogut-Susskind (KS) fermion. List of implemented algorithms:
\begin{itemize}
\item multiplication by $Q$, for unimproved and $O(a^2)$ 
(Asqtad) improved actions~\cite{lepage}
\item BiCGStab inversion for $Q$
\item BiCGStab inversion for $Q$ using the UML decomposition~\cite{milc}
\item function {\tt make\_meson}
\end{itemize}
These algorithms work for arbitrary gauge groups $SU(N_c)$ and for
an arbitrary even number of dimensions (except {\tt make\_meson}). 
The multiplication by $Q$, both improved and unimproved, for $SU(3)$, 
is optimized using Pentium 4 SSE2 instructions in assembler language.
In the unimproved case only half of the SSE2 registries are used and there 
is room for an extra factor two in speed.
The function {\tt make\_meson} builds any meson propagator 
(made out of staggered quarks) for arbitrary Spin$\otimes$Flavour 
structure. This algorithm is described in 
ref.~\cite{mesons}
\vskip 3mm
\noindent{\bf class {\tt staggered\_propagator}:}

This is an implementations of the staggered  
propagator consisting of 16 sources contained in a $2^4$ hypercube at 
the origin of the lattice. A {\tt staggered\_propagator} can be used to 
propagate any hadron from the hypercube at the origin of the lattice to 
any other hypercube without extra inversions.
\vskip 3mm

All fields in FermiQCD inherit the standard I/O methods of MDP ({\tt save} and {\tt load}) and the file format is independent on the lattice 
partitioning over the parallel processes. These I/O functions, as well as 
all the FermiQCD algorithms, are designed to optimize 
interprocess communications.

\section{Example}
We present here, as an example, a full program that generates 100 $SU(3)$ 
gauge configurations ({\tt U}), starting from a hot one. On each configuration
it computes a pion propagator ({\tt pion}) made of $O(a^2)$ improved 
quark propagators and prints it out.
These propagators are computed using the SSE2 optimized clover action 
and the BiCGStab inversion algorithm. The program works in parallel.
\newpage
\hrule height 1pt
\begin{verbatim}
#define PARALLEL
#include "fermiqcd.h"

void main(int argc, char **argv) {
   mpi.open_wormholes(argc, argv);

   int t,a,b,conf;
   int nc=3, box[4]={16,8,8,8};
   generic_lattice  L(4,box);
   gauge_field      U(L,nc);
   fermi_propagator S(L,nc);
   site             x(L);
   float            pion[16];
   U.param.beta=5.7;
   S.param.kappa=0.1345;
   S.param.cSW=1.5;

   default_fermi_action=
                       mul_Q_Luscher;
   default_inversion_method=
                  BiCGStab_inversion;

   set_hot(U);
   heatbath(U,100);
   for(conf=0; conf<100; conf++) {
      heatbath(U,30);
      compute_em_field(U);
      generate(S,U);
      for(t=0; t<16; t++) pion[t]=0;
      forallsites(x)
         for(a=0; a<4; a++)
            for(b=0; b<4; b++)
               pion[x(TIME)]+=
                  real(trace(S(x,a,b)*
                  hermitian(S(x,b,a))));
      mpi.add(pion, 16);
      if(ME==0) for(t=0; t<16; t++)
         printf("%i %f\n", t, pion[t]);
   }
   mpi.close_wormholes();
}
\end{verbatim}
\hrule height 1pt \vskip 2mm

\begin{figure}
\begin{center}
\epsfxsize=7.5cm
\epsfysize=6cm
\epsfbox{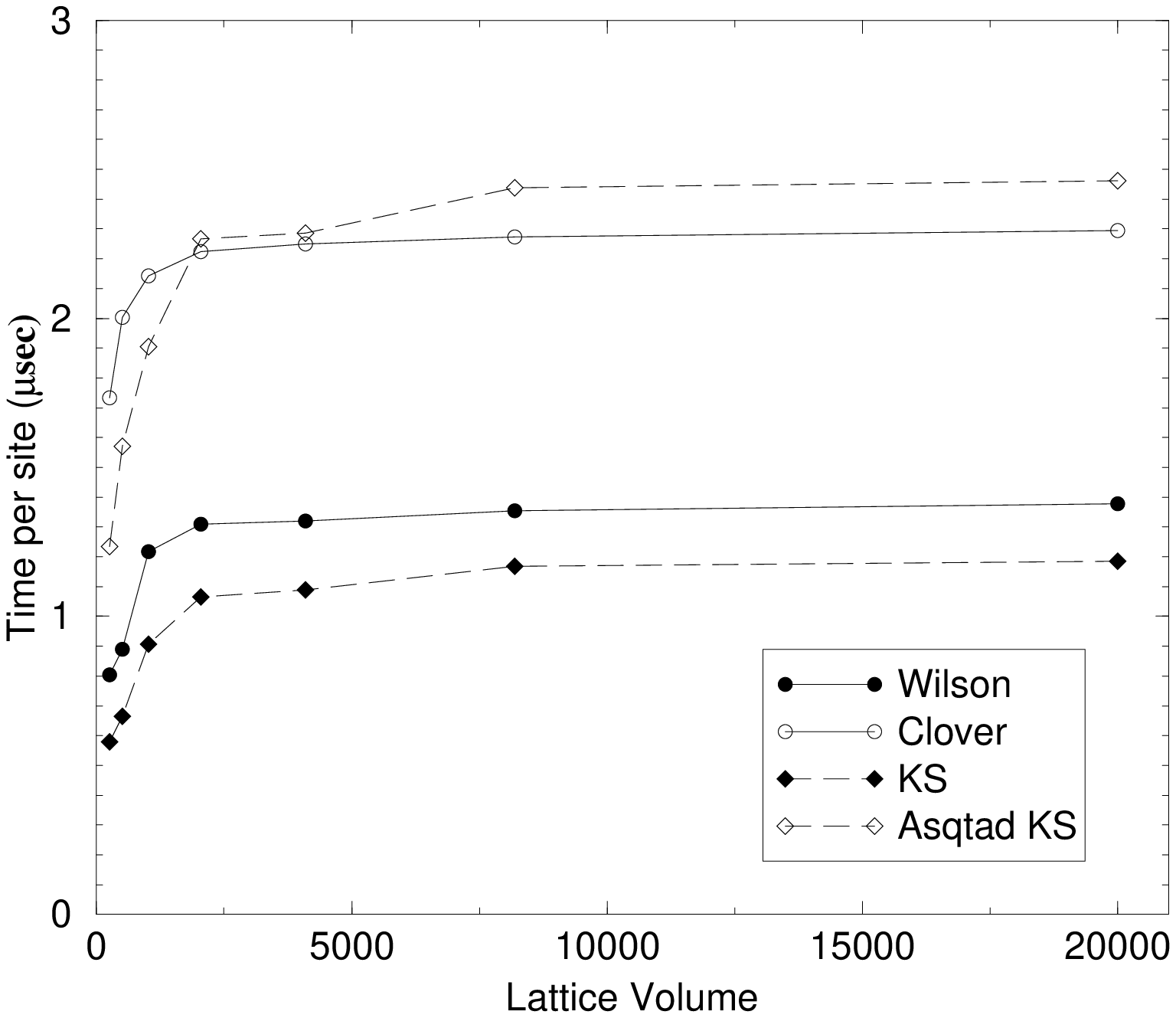}
\vskip -1cm
\caption{Time per site in $\mu$sec for {\tt mul\_Q\_Luscher} (Wilson, clover, KS and Asqtad KS in single precision).\label{bench1}}
\end{center}
\end{figure}

Comments:
\begin{itemize}
\item {\tt L} is the user-defined name of the lattice ($16\times8^3$)
\item {\tt U}, {\tt S} and {\tt x} are the gauge field, the fermi propagator and an auxiliary {\tt site} variable defined on the lattice {\tt L}
\item {\tt default\_fermi\_action} is a pointer to the function
that implements the action to be used. {\tt mul\_Q\_Luscher} is one of the
the built-in clover actions, optimized for Pentium 4.
\item {\tt default\_inversion\_method} is a pointer to the function 
that implements the inversion algorithm (minimal residue or BiCGStab)
\item {\tt compute\_em\_field} computes the chromo-electro-magnetic field 
required by the action\footnote{%
The chromo-electro-magnetic field is a member variable of the gauge field.
FermiQCD has almost no global variables except pointers to the 
functions that implement the algorithms.}.
\item {\tt generate} computes the quark propagtor {\tt S} on the 
given gauge configuration {\tt U}.
\item {\tt forallsites(x)} is a parallel loop on {\tt x}. 
Each processor loops on the local sites.
\item {\tt mpi.add(pion,16)} sums the vector {\tt pion[16]} in parallel.
\item {\tt if(ME==0)} guarantees that only one processor performs the 
output.
\end{itemize}

\section{Benchmarks}

\begin{figure}
\begin{center}
\epsfxsize=7.5cm
\epsfysize=6cm
\epsfbox{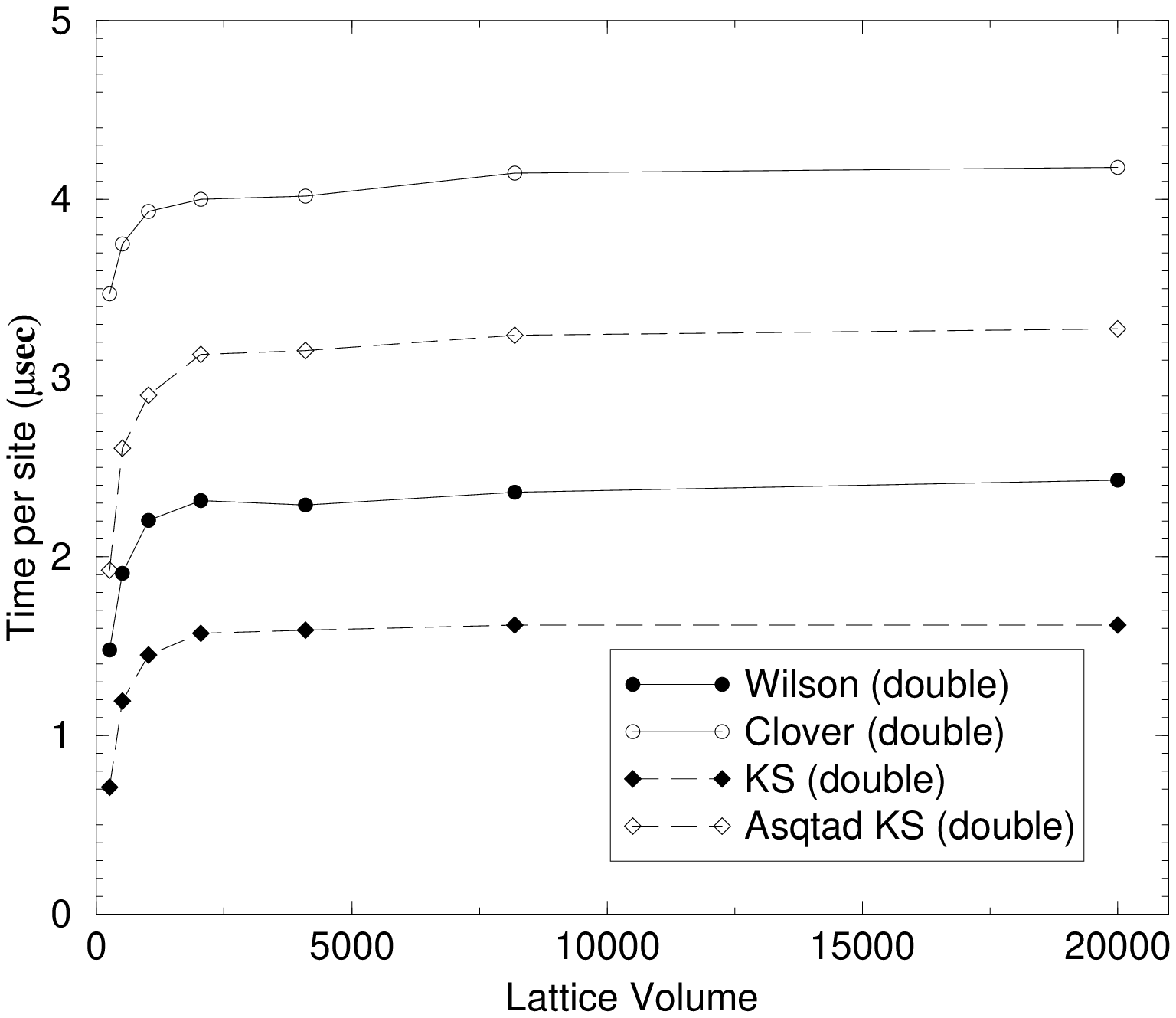}
\vskip -1cm
\caption{Time per site in $\mu$sec for {\tt mul\_Q\_Luscher} (Wilson, clover, KS and Asqtad KS in single precision).\label{bench2}}
\end{center}
\end{figure}

In fig.~\ref{bench1} and fig.~\ref{bench2} we report some benchmarks 
for the multiplication by $Q$ of a fermionic field, for the different 
actions (using a single CPU Pentium 4 PC running at 1.4 GHz,
Linux 2.4 and gcc 2.95.3).
\vskip 2mm
This work was performed at Fermilab (U.S. Department of Energy Lab 
(operated by the University Research Association, Inc.), under 
contract DE-AC02-76CHO3000.


\begin{thebibliography}{999}

\bibitem{lattice} M.~Di~Pierro, hep-lat/0009001. (Lattice QCD tutorial 
with examples in FermiQCD)
\bibitem{mdp} M.~Di~Pierro, hep-lat/0004007. (Updated tutorial on MDP)
To be published on CPC.  
\bibitem{luscher} M.~L\"uscher, {\it these proceedings}
\bibitem{lepage} G.~P.~Lepage, Phys.Rev. D59 (1999) 074502
\bibitem{milc} K.~Orginos {\it at al.} (MILC), Phys.Rev. D59 (1999) 014501
\bibitem{mesons} M.~Di~Pierro, {\it in preparation}
\end{thebibliography}
\end{document}